\documentclass[sn-mathphys]{sn-jnl}

\jyear{2025}%

\theoremstyle{thmstyleone}%
\theoremstyle{thmstyletwo}%

\theoremstyle{thmstylethree}%

\begin{document}

\title[EPR, Bell and Quantum Correlations]
{EPR Paradox, Bell Inequalities and Peculiarities of Quantum Correlations}


\author*[1,2]{\fnm{Apoorva D.} \sur{Patel}}\email{adpatel@iisc.ac.in}

\affil[1]{\orgdiv{Centre for High Energy Physics}, \orgname{Indian Institute of Science}, \orgaddress{\city{Bengaluru} \postcode{560012}, \state{Karnataka}, \country{India}}}
\affil[2]{\orgdiv{International Centre for Theoretical Sciences}, \orgname{TIFR}, \orgaddress{\city{Bengaluru} \postcode{560089}, \state{Karnataka}, \country{India}}}


\abstract{Quantum theory revolutionised physics by introducing a new
fundamental constant and a new mathematical framework to describe the
observed phenomena at the atomic scale. These new concepts run counter
to our familiar notions of classical physics, and pose questions about
how to understand quantum physics as a fundamental theory of nature.
Peculiarities of quantum correlations underlie all these questions,
and this article describes their formulation, tests and resolution
within the standard framework of quantum theory.}


\maketitle

In the early years of twentieth century, properties of photons and atomic
energy spectra challenged the existing framework of classical physics.
To explain the observations, a new fundamental constant had to be introduced,
the Planck's constant. Even with its incorporation, semi-classical physics
could not fully describe the observed phenomena, and a completely new
theoretical framework had to be constructed, named quantum physics.
This new framework brought in non-commuting observables and probabilistic
events, which raised many questions about its interpretation that still
continue to be debated.

Section 1 of this review introduces the foundational debate between what
is real and what is observable, and follows it up with the Bell inequality
based on locality and the Leggett-Garg inequality based on macrorealism.
Section 2 describes the experimental tests that violate these inequalities
and confirm the standard quantum mechanical analysis. Section 3 covers
various scenarios that have been proposed for the interpretation of the
observed but non-intuitive quantum phenomena. Section 4 describes other
peculiar features of quantum correlations: non-contextuality, the Wigner
function formulation, the Schmidt decomposition and quantum statistics.
Section 5 presents a short outlook for this subject.

\section{Interpreting Quantum Mechanics}\label{sec1}

The question of interpretation of quantum mechanics goes all the way back to
its origin. Even though Einstein contributed to many early developments in
quantum mechanics, he was uncomfortable with its probability interpretation.
He was not satisfied with quantum mechanics being treated as an empirical
theory; he wanted quantum mechanics to arise from a deterministic underlying
structure, similar to how macroscopic statistical mechanics arises from
microscopic atomic scale phenomena. The Einstein-Podolsky-Rosen (EPR) paper
posed this question directly: Can quantum-mechanical description of physical
reality be considered complete? \cite{EPRpair}

Bohr responded to the EPR paper, in the same journal, with the same title.
He reiterated his Copenhagen interpretation, and that did not attract much
attention. Schr\"odinger responded as well, sharpening Einstein's question.
That response is remembered well for the two concepts he introduced. One
is that of ``entanglement", i.e. unusual quantum correlations between two
separated parts of a system \cite{SchEntangle}. The other is that of a
``cat" (named after him), which could be dead or alive depending on the
occurrence of a quantum event \cite{SchCat}. The philosophical debate on
these peculiarities, often referred to as the ``hidden variable" problem,
still goes on.

Subsequently, Bohm rephrased the question of quantum correlations in the
setting of a finite dimensional system \cite{BohmPair}, which turned out to
be crucial for performing accurate experimental tests. For this setting,
Bell showed that the observable correlations must obey an inequality,
when the hidden variables of quantum theory satisfy certain properties
\cite{BellIneq}. This analysis has been extended to different quantum
systems and different observable correlations, and has generated a lot of
discussion about interpretation of quantum mechanics \cite{BellBook}.
The 2022 Nobel Prize in Physics was awarded to John F. Clauser, Alain Aspect
and Anton Zeilinger, for performing accurate experiments that demonstrated
that the Bell inequality is clearly violated by quantum correlations between
two photons produced in a singlet state. The Nobel citation reads ``for
experiments with entangled photons, establishing the violation of Bell
inequalities and pioneering quantum information science" \cite{PhyNobel22}.

\subsection{The Fundamental Conundrum}

At the heart of the interpretation of quantum mechanics is a quandary
described by two Greek words, ontology and epistemology. The former
concerns finding out what is real, irrespective of the observers.
The latter focuses on what is observable in practice, and that may
depend on the capability of the observer.

It is well-established that quantum dynamics produces probabilistic outcomes,
and the measurement postulate of quantum mechanics successfully gives the
prescription to predict the probability distribution. But what has remained
mysterious is how and why the probabilistic outcomes arise, and whether the
observer plays any role in that. Probabilistic description of physical
phenomena is routine in statistical physics. It is understood as arising
from an ensemble of underlying dynamics, which is unobserved and hence
summed (or integrated) over all possibilities that may occur. The mystery
then is: Can the quantum indeterminacy be explained as arising from so far
unobserved ``hidden variables"?

The use of ``effective theories", valid within specific ranges of their
degrees of freedom, is widespread in physics. Such theories provide an
excellent description of the observed data, in terms of certain empirically
adjusted parameters. These parameters are understood to be consequences of
the unobserved degrees of freedom (apart from fundamental constants), and
carry information about their dynamics. For example, a fluid is generally
described as a continuous medium, while its properties such as temperature,
density and pressure parametrise the underlying atomic dynamics. Moreover,
the underlying atomic dynamics produces observable signals in certain
correlations, such as the Brownian motion of a particle in a fluid and
the fluctuation-dissipation relation.

The hidden variables of quantum mechanics must have a distribution to produce
probabilistic outcomes. Even when they are integrated out, they would leave
behind observable parameters and contributions to correlations. The question
then is whether we can learn something about the properties of the hidden
variables by observing their consequences in the effective description.
It is in this sense that the peculiarities of quantum correlations takes
the centre-stage in trying to figure out the nature of the hidden variables.

It should be noted that physical parameters and correlations arising from
global conservation laws do not conflict with the locality of relativity.
They represent certain symmetries of the overall dynamics, and are part of
inherent features of nature. For example, when a firecracker bursts and
one half of it is then found at one place, it can be immediately inferred
that the other half went in the opposite direction (as dictated by the
conservation of momentum) without making a separate observation or worrying
about instantaneous communication of information. The peculiarities of
quantum correlations go beyond such situations, and that was emphatically
illustrated by Bell.

\subsection{Bell Inequality \cite{BellIneq}}\label{sec1.2}

The quantum correlations pointed out by EPR, and rephrased by Bohm, concern
two-particle singlet states, created at a common origin and then evolved so
that the two components appear at a space-like separation. In case of photons,
such situations arise in the two-photon cascade transitions of certain atoms,
or the two-photon decay of a neutral pion, where both the initial and the
final states have zero total momentum and zero total angular momentum. The 
zero momentum implies that the two photons fly off in opposite directions,
while the zero angular momentum implies that the internal states of the two
photons are anti-correlated, in terms of their spins or polarisations. The
quantum mechanical description of this entangled singlet internal state is:
$\vert\psi_{12}\rangle = \frac{1}{\sqrt{2}}
(\vert\uparrow\downarrow\rangle-\vert\downarrow\uparrow\rangle)$.
If only one of the photons is observed, it is found to be in either
$\vert\uparrow\rangle$ or $\vert\downarrow\rangle$ state with equal
probability.

Now consider the situation where two space-like separated observers A and B
measure the two photon spins along non-parallel and non-orthogonal directions,
say $\vec{a}$ and $\vec{b}$. The measurement outcomes are then probabilistic,
and let us label them as $A(\vec{a}),B(\vec{b})\in\{\pm1\}$. The directions
are chosen such that conservation of angular momentum offers no relation
between $A(\vec{a})$ and $B(\vec{b})$, and the property to be investigated
is the correlation between the two.

Next, since the two photons have a common origin, let us imagine that the
distributions of $A(\vec{a})$ and $B(\vec{b})$ arise from some common
ensemble of underlying hidden variables $\{\lambda\}$. The hidden variables
appear at the point of origin of the two photons, and are then carried by
the photons till the points of their observation. We assign to the ensemble
of hidden variables a normalised weight distribution $\rho(\lambda)$, with
$\int d\lambda~\rho(\lambda)=1$, and relabel the measurement outcomes as
$A(\vec{a},\lambda)$ and $B(\vec{b},\lambda)$ to express their implicit
dependence on the hidden variables.

The two-point correlation of the measurement outcomes in this setting is:
$P(\vec{a},\vec{b})
= \int d\lambda~\rho(\lambda)~A(\vec{a},\lambda)~B(\vec{b},\lambda)$,
and global spin conservation implies $B(\vec{b},\lambda)=-A(\vec{b},\lambda)$.
Then, using the property that $A(\vec{b},\lambda)^2=1$, we can construct the
following difference of correlations:
\begin{equation}
P(\vec{a},\vec{c})-P(\vec{a},\vec{b})
= \int d\lambda~\rho(\lambda)~A(\vec{a},\lambda)~A(\vec{b},\lambda)
~[1-A(\vec{b},\lambda)~A(\vec{c},\lambda)] ~.
\end{equation}
This difference obeys a simple bound, following from the triangle inequality
$\vert x+y\vert \le \vert x\vert+\vert y\vert$ (the sum can be replaced by
an integral),
\begin{align}
\vert P(\vec{a},\vec{c})-P(\vec{a},\vec{b}) \vert
& \le \int d\lambda~\vert \underbrace{\rho(\lambda)}_{\ge0} \vert
~\underbrace{\vert A(\vec{a},\lambda)~A(\vec{b},\lambda) \vert}_{=1}
~\vert \underbrace{[1-A(\vec{b},\lambda)~A(\vec{c},\lambda)]}_{\ge0} \vert
\nonumber \\
& = 1+P(\vec{b},\vec{c}) ~.
\label{bellineq}
\end{align}
Here the simplification uses the properties indicated below the equation;
the last factor of the integrand is non-negative and so the absolute value
sign is dropped, the middle factor of the integrand is dropped since it is
equal to one, and the absolute value sign of the first factor is dropped
assuming that the ensemble weight of the hidden variables is non-negative.

In quantum mechanics, the spin (or polarisation) operator producing the
measurement outcome $A(\vec{a},\lambda)$ is $\vec{\sigma}_1\cdot\vec{a}
\equiv (\sigma_1)_xa_x + (\sigma_1)_ya_y + (\sigma_1)_za_z$.
The two-point correlation is then $P(\vec{a},\vec{b})
= \langle(\vec{\sigma}_1\cdot\vec{a})(\vec{\sigma}_2\cdot\vec{b})\rangle
= -\vec{a}\cdot\vec{b}$,
due to anti-correlation of the spin components $(\sigma_1)_i$ and
$(\sigma_2)_i$. This correlation violates the bound derived above in 
Eq.\eqref{bellineq} for many choices of $\vec{a},\vec{b},\vec{c}$.
For example, choosing the
directions $\vec{a}=\uparrow$, $\vec{b}=\nearrow$, $\vec{c}=\rightarrow$
in two-dimensional space yields,
$P(\vec{a},\vec{b}) = -\frac{1}{\sqrt{2}} = P(\vec{b},\vec{c})$
and $P(\vec{a},\vec{c}) = 0$, while
$\vert0+\frac{1}{\sqrt{2}}\vert \not\leq 1-\frac{1}{\sqrt{2}}$.
 
\subsection{Leggett-Garg Inequality \cite{LeggettGarg}}

The Bell inequality constrains correlations between two halves of a quantum
system at space-like separation. Leggett and Garg formulated a different
constraint for correlations between states of a single quantum system
at different times. In their set up, a classical macrorealistic system
possesses three properties: the system is in one of its distinct possible
states at every instance, this state can be determined with negligible
perturbation to the subsequent evolution, and any measurement result cannot
be affected by what is measured later. They demonstrated that an inequality
following from macrorealism is violated by the non-classical nature of
temporal correlations in a quantum system. In a sense, the assumption of
non-invasive measurements in macrorealism replaces that of local measurements
in Bell's analysis.

The simplest example concerns the state of a two-state quantum system (qubit)
at three different times. Let $Q(t_i) \in \{\pm1\}$ be the binary observable
measured at each time, and construct the correlations
$C_{ij} = \langle Q(t_i)Q(t_j) \rangle$ for all $t_i>t_j$. Then, macrorealism
implies that the combination $K_3 \equiv C_{21}+C_{32}-C_{31} \in [-3,1]$.
This Leggett-Garg inequality follows from just enumerating the eight
possibilities for $Q(t_i)$, and observing that for each of them $K_3$ is
either $1$ or $-3$. Any probabilistic combination of the eight possibilities
would therefore keep $K_3 \in [-3,1]$. Note that it is not necessary to
specify the initial state of the quantum system.

For a qubit, we can choose $\widehat{Q}(t_i)=\vec{a}_i\cdot\vec{\sigma}$
with unit vectors $\vec{a}_i$. Then, as per the Born rule where projective
measurements alter the quantum state, the observed correlations are:
\begin{equation}
C_{ij} = \langle\vec{a}_i\cdot\vec{\sigma} ~ \vec{a}_j\cdot\vec{\sigma}\rangle
       = \vec{a}_i\cdot\vec{a}_j ~.
\end{equation}
Let $\theta_{ij}$ be the angle between $\vec{a}_i$ and $\vec{a}_j$.
Choosing all $a_i$ to be coplanar with $\theta_{21}=\theta_{32}=\pi/3$ and
$\theta_{31}=2\pi/3$, gives $K_3$ its maximum value, $K_3^{\rm max}=3/2$,
which violates the Leggett-Garg inequality. (Note that always $K_3\ge-3$.)

The four-term version of the Leggett-Garg inequality, $K_4 \in [-2,2]$ for
$K_4 \equiv C_{21}+C_{32}+C_{43}-C_{41}$, is related to the CHSH inequality
of Eq.\eqref{CHSHbound}.

\section{Experimental Tests}

\subsection{Bell Test Experiments}

Bell's derivation of the two-photon correlation inequality provided a clear
target for experimentalists to test. Of course, they had to develop the
technology that would produce reliable singlet photon-pair sources, detect
single photons with high success rate and measure their polarisations to
high accuracy. They also had to close many loopholes, so that the observed
correlations connect to the hidden variable properties and not to other
extraneous coincidences. These developments occurred in several stages.

Compared to the correlation check described by Bell, a modified version
proposed by Clauser-Horne-Shimony-Holt (CHSH) is easier to implement
experimentally \cite{CHSHineq}. In this version, A chooses one of two
polarisation measurement directions differing by angle $\frac{\pi}{4}$,
B does the same, and and B's measurement directions are rotated from those
of A by angle $\frac{\pi}{8}$. A linear combination of the the four possible
polarisation correlations (labeled by the measurement directions
$\vec{a},\vec{a}',\vec{b},\vec{b}'$) then satisfies a Bell-type inequality:
\begin{equation}
\vert P(\vec{a},\vec{b}) + P(\vec{a},\vec{b}')
+ P(\vec{a}',\vec{b}) - P(\vec{a}',\vec{b}') \vert \le 2 ~.
\label{CHSHbound}
\end{equation}
This CHSH inequality follows from the observation that for every assignment
of the measurement outcomes
$\{A(\vec{a}),A(\vec{a}'),B(\vec{b}),B(\vec{b}')\} \in \{\pm1\}$,
the left hand side of the above formula evaluates to $\pm2$.

Having contributed to the CHSH proposal as a graduate student, Clauser took
up the challenge to test the inequality as a postdoctoral fellow at Berkeley.
He did not have research funds. So the experiment was carried out using
borrowed equipment and some discarded parts in a basement laboratory,
together with graduate student Freedman. Calcium atoms were used to generate
entangled photon pairs, and the four correlation terms of the CHSH inequality
were measured one by one. The observed correlation results clearly violated
the inequality, asserting the peculiar nature of quantum correlations.

The Clauser-Freedman experiment did not test the assumption made by Bell
that there is no communication of any information between the measurements
performed by A and B. Aspect and his collaborators at Orsay overcame this
shortcoming by refining the experiment. The entangled photon pairs were
generated at a higher rate, and the polarisation measurement directions on
either side were randomly switched at a rate faster than the time light
took to travel between A and B. The observed violation of the CHSH inequality
was stronger and in accordance with the quantum mechanical prediction.

Zeilinger and his collaborators at Innsbruck and Vienna later conducted more
refined tests of Bell-type inequalities, to firmly close the communication
loophole. For CHSH inequality tests, entangled photon pairs were created by
shining a laser on a special crystal, and random numbers switching between
polarisation measurement directions were constructed using signals from
distant galaxies to rule out any bias. They also demonstrated quantum
teleportation using a two-photon entangled state, where the quantum state
of a photon disappears from one location and reappears at a distant location
without any material transfer. This demonstration was then extended to
entanglement swapping, creating a quantum entangled state between two
parties who have not interacted in the past.

\subsection{GHZ Test \cite{GHZtest}}

Greenberger, Horne and Zeilinger considered quantum correlations beyond the
two-qubit correlations of Bell states. They proposed a correlated three-qubit
state, which provides a deterministic separation (in contrast to probabilistic
expectation values) between answers predicted by a Bell-type analysis and
physical quantum measurement. This GHZ state is
$\frac{1}{\sqrt{2}}(\langle 000\vert - \langle 111\vert)$.
It is a simultaneous eigenstate with eigenvalue $1$ of the three commuting
operators:
$A^{(1)}B^{(2)}B^{(3)}$, $B^{(1)}A^{(2)}B^{(3)}$ and $B^{(1)}B^{(2)}A^{(3)}$,
with $A^{(i)}=\sigma_1^{(i)}$ and $B^{(j)}=\sigma_2^{(j)}$, where the
superscript denotes the qubit position. It is also an eigenstate with
eigenvalue $-1$ of the fourth operator $A^{(1)}A^{(2)}A^{(3)}$.

All the operators $A^{(i)}$ and $B^{(j)}$ have eigenvalues $\pm1$, and they
square to identity. The three factors of the three-qubit operators listed
above can be independently measured in experiments, and multiplied to obtain
the operator eigenvalues. Also, all four three-qubit operators commute and
their eigenvalues can be measured in any order. Classically, with these
properties, all the measurement values can be multiplied, and the product
of all the factors in the first three operators gives the factors of the
fourth operator (all $B^{(j)}$ factors get squared). Then the fourth operator
should give the eigenvalue $1$, contrary to the quantum eigenvalue $-1$.

The reason for the difference between classical and quantum predictions is
that, with anticommutation of $A^{(i)}$ and $B^{(i)}$, the product of the
first three operators gives minus the fourth operator:
\begin{align}
& (A^{(1)}\otimes B^{(2)}\otimes B^{(3)})
~ (B^{(1)}\otimes A^{(2)}\otimes B^{(3)})
~ (B^{(1)}\otimes B^{(2)}\otimes A^{(3)}) \\
= & (A^{(1)}B^{(1)}B^{(1)}) \otimes (B^{(2)}A^{(2)}B^{(2)}) \otimes
(B^{(3)}B^{(3)}A^{(3)}) 
= - A^{(1)} \otimes A^{(2)} \otimes A^{(3)} ~. \nonumber
\end{align}
This GHZ example thus connects the non-commutative nature of quantum mechanics
to a deterministic outcome.

Experimentally, the GHZ states can be prepared probabilistically, purified
sufficiently by suitable measurements, and then the eigenvalue of the fourth
operator can be measured. Zeilinger and his group performed such a test, and
it showed that indeed the quantum mechanical prediction is correct.

\subsection{Leggett-Garg Inequality Tests}

The Leggett-Garg macrorealistic inequality can be applied to systems of
arbitrary size. Indeed, one of the motivations for developing it was to
check whether macroscopic objects can exist in a superposition of states.
Its tests have been performed for microscopic objects such as photons or
nuclear spins, for larger size superconducting transmon systems, and in
a non-invasive setting that can be size independent with a two-state
ancilla as the measuring apparatus \cite{LGtest}.

An ideal non-invasive binary measurement of a system can be performed by a
two-state ancilla such that one of the states of the system flips the ancilla
while the other one does nothing (this is known as the C-NOT operation).
Then in the post-selected subset of an ensemble of experiments where the
ancilla is unaffected, the system state is determined non-invasively and its
subsequent evolution can be investigated. How the initial state of the system
or the ancilla was prepared does not matter in this scenario.

This strategy can be used to evaluate the Leggett-Garg correlation $C_{ij}$.
For non-invasive measurements of $Q(t_j)$, two ensembles of experiments are
put together---the first system state is non-invasively post-selected in one
and the second system state is non-invasively post-selected in the other
(with a complementary set-up). No post-selection is necessary after
measurements of $Q(t_j)$, because the system evolution beyond $t_j$ is of
no consequence. Thus the combination $K_3$ is evaluated using six separate
ensembles of experiments. The expectation value determinations can be
statistically improved, by treating multiple impurities embedded in a
material as an ensemble, provided that the impurities are dilute enough
to interact negligibly with each-other.

Such a test, carried out using an ensemble of nucleus-electron spin pairs
in phosphorus-doped silicon, confirms the quantum prediction, disproving
macrorealism \cite{LGtest}. In particular, the experimental parameters
could attain the values (i.e. sufficiently low temperature in a stable
magnetic resonance set-up), needed for accurate ancilla preparation as
well as for negligible perturbation to the controlling system state during
the non-invasive measurement.

\section{The Way Out}

Experimental tests of all Bell-type inequalities demonstrate that observed
physical correlations between photons (or spins) agree with the standard
quantum mechanical analysis, and disagree with constraints derived using
hidden variables with certain properties. These repeated confirmations of
quantum mechanical predictions reiterate the fundamental question: What is
the origin of the peculiar quantum correlations? Obviously, at least one of
the assumptions in the derivation of inequalities must be given up. We do
not want to give up the conservation laws or causality, because that would
destroy the framework of physics at its core. Also, quantum dynamics and
special relativity have been successfully merged in quantum field theory
and verified to a fantastic level of accuracy. So we know that there is no
need to give up one or the other. We need to therefore inspect more subtle
ingredients in the analysis to find a credible interpretation of quantum
mechanics.

{\bf Give up locality:}
This is the frequently used label, i.e. quantum mechanics is non-local.
(It does not mean the same thing as existence of non-local quantum 
correlations, which the experiments verify.) Some of the hidden variables
in this case are non-local and unobservable (to avoid conflict with special
relativity), and the de Broglie-Bohm theory is an explicit example of this
scenario.

{\bf Give up statistical independence:}
This is a difficult to overcome loophole. In this case, the hidden variables
in some way depend on the measurement settings, so the observables are
influenced by the measurement apparatus. Superdeterministic, retrocausal
and supermeasured theories with such properties have been constructed,
keeping in mind the fact that correlation is not the same as causation.

{\bf Give up positivity:}
This is how the standard formulation of quantum mechanics works, without
giving up locality or statistical independence. The quantum ensemble weights
are allowed to be negative, i.e. $\rho(\lambda)\not\geq0$. Indeed, the
quantum density matrix is such a weight, with $Tr(\rho O)$ providing the
expectation value of a physical observable $O$. The quantum density matrix
is a Hermitian generalisation of the classical probability distribution,
and its off-diagonal elements contribute to the quantum correlations tested
with non-parallel and non-orthogonal directions. In the Wigner function form
\cite{wigner}, written down before the EPR paper, the quantum density matrix
becomes real. Physically observable quantities require smearing the Wigner
function over an area in the phase space (with the characteristic scale
$\Delta x\Delta p\sim h$), which wipes out locally negative weights and
restores positivity of observed probabilities. Section \ref{sec4.2}
discusses this attribute of the quantum density matrix in more detail.
How such a density matrix may arise in quantum mechanics, and the dynamics
of what really happens during quantum measurement, remain open questions
for a different level of analysis.

It must be emphasised that negative weights in the analysis of a physical
problem are not an obstacle of principle. As an example, consider the
diffusion equation describing evolution of temperature over a region.
It is routinely solved by decomposing the temperature into its Fourier
eigenmodes, and then determining the contribution of each eigenmode.
By definition, the Fourier eigenmodes are sinusoidal functions, giving both
positive and negative contributions. On the other hand, the temperature
that is the sum of all Fourier eigenmodes is always positive. There is no
conflict of any kind, and the important lesson is that physical reality
should not be demanded of mathematical variables.

\section{Quantum Correlations}

The previous sections discussed topics that arose from an effort to
interpret quantum mechanics as an effective theory, and the constraints
it imposes on the hidden variables. Although some philosophical questions
remain, all the experimental tests have emphatically confirmed the standard
formulation of quantum theory. There are other peculiarities of quantum
correlations that mathematically follow from this standard formulation of
quantum mechanics, and we describe them in this section.

\subsection{Contextuality \cite{mermin}}

The Bell-Kochen-Specker theorem uses the non-commuting operator structure
of quantum mechanics in a clever way to elucidate the property that the
results of quantum measurements are context-dependent and do not just
reproduce a pre-existing value \cite{bellks1,bellks2}. Moreover, it is not
necessary to specify the quantum state on which this property can be tested.

Consider a system of two qubits, for which certain Pauli operators are
measured. Pauli operators for the same qubit mutually anticommute, while
those for different qubits commute. Also, the eigenvalues for all Pauli
operators are $\pm1$. Now consider measurement of the nine Pauli operators
that are arranged in a $3\times3$ array as follows:
\begin{center}
\begin{tabular}{ccc}
$\sigma_x^{(1)}I^{(2)}$ & $I^{(1)}\sigma_x^{(2)}$ & $\sigma_x^{(1)} \sigma_x^{(2)}$ \\
$I^{(1)}\sigma_y^{(2)}$ & $\sigma_y^{(1)}I^{(2)}$ & $\sigma_y^{(1)} \sigma_y^{(2)}$ \\
$\sigma_x^{(1)} \sigma_y^{(2)}$ & $\sigma_y^{(1)} \sigma_x^{(2)}$ & $\sigma_z^{(1)} \sigma_z^{(2)}$ \\
\end{tabular}
\end{center}
This arrangement is such that the three Pauli operators in each row and each
column mutually commute, and so can be simultaneously measured. The product
of the three operators in each row and the first two columns is $+1$, while
the product of the three operators in the third column is $-1$. Hence, the
product of all nine observables is $+1$ row-wise, but $-1$ column-wise, and
that is impossible to satisfy. We can only surmise that measurements of
the operators on different rows (or different columns) require different
experimental arrangements, and there is no a priori reason to believe that
these arrangements would not affect the results. In other words, the results
of observations do not depend just on the state of the system but also on
the full set-up of the apparatus.

This example illustrates that in quantum mechanics the value of an observable
when it is part of one mutually commuting set may not be the same when it is
part of another mutually commuting set, when some of the members of one set
do not commute with some of the members of the other. This property is called
``contextuality". Subsequent to this deduction, Bell considered spatially
separated observations of two qubits that guarantee non-contextuality as per
relativity, and derived his inequality based on the assumption of locality.
This is already discussed in Section \ref{sec1.2}.

\subsection{Wigner Function}\label{sec4.2}

The quantum density matrix generalises the concept of classical probability
distribution, and encodes complete information about a quantum system.
Expectation value of any observable can be calculated from it by the
simple rule, $\langle O\rangle = Tr(\rho O)$. The Wigner function is the
quantum density matrix, in the representation where its relative index is
Fourier transformed to its conjugate variable \cite{wigner}. It is real by
construction. Since it is defined in the symplectic phase space, its domain
is quantised in units of the Planck constant.

\subsubsection{Infinite dimensional systems}

The Wigner function for a continuous one-dimensional quantum state is:
\begin{eqnarray}
W(x,p) &=& \frac{1}{2\pi\hbar} \int_{-\infty}^{\infty} dy ~
           \psi^*(x-\frac{y}{2}) e^{ipy/\hbar} \psi(x+\frac{y}{2}) \nonumber\\
       &=& \frac{1}{2\pi\hbar} \int_{-\infty}^{\infty} dy ~
           \rho(x-\frac{y}{2},x+\frac{y}{2}) e^{ipy/\hbar} ~,
\end{eqnarray}
\begin{equation}
\rho(x-\frac{y}{2},x+\frac{y}{2}) = \int_{-\infty}^{\infty}
                                    dp~W(x,p) e^{-ipy/\hbar} ~.
\end{equation}
It can be negative, but its marginals are non-negative.
\begin{equation}
\int_{-\infty}^{\infty} dp~W(x,p) = \vert\psi(x)\vert^2 = \rho(x,x) ~,~~
\int_{-\infty}^{\infty} dx~W(x,p) = \vert\tilde{\psi}(p)\vert^2 ~.
\end{equation}
Its smeared values over a phase space volume element
$\Delta x \Delta p=2\pi\hbar$ (associated with counting
of states in quantum statistics) are also non-negative.
The normalisation condition is:
\begin{equation}
{\rm Tr}(\rho)=1 \quad\longleftrightarrow\quad
\int_{-\infty}^{\infty} dx~dp~W(x,p) = 1 ~.
\end{equation}

The expectation value of a Hermitian operator $O$ is obtained as:
\begin{eqnarray}
\langle O\rangle \equiv {\rm Tr}(\rho O)
&=& \int dx~dy~\rho(x-\frac{y}{2},x+\frac{y}{2})~O(x+\frac{y}{2},x-\frac{y}{2}) \nonumber\\
&=& \int dx~dy \int dp~W(x,p)e^{-ipy/\hbar} \int dq~O(x,q)e^{iqy/\hbar} \nonumber\\
&=& 2\pi\hbar \int dx \int dp~W(x,p) \int dq~O(x,q)~\delta(p-q) \nonumber\\
&=& 2\pi\hbar \int dx~dp~W(x,p)~O(x,p) ~.
\end{eqnarray}
It should be noted that $O(x,p)$ implicitly defined here is Hermitian,
and its normalisation is fixed by the convention $\langle I\rangle=1$.

\subsubsection{Finite dimensional systems}

For a finite dimensional quantum system with $d$ degrees of freedom,
the odd and even values of $d$ need to be handled separately.
When $d$ is odd,
\begin{equation}
W(n,k) = \frac{1}{d} \sum_{m=0}^{d-1} \rho_{n-m,n+m} e^{4\pi ikm/d} ~,
\end{equation}
is a valid Wigner function \cite{dimodd,primefac}. Here the indices are
defined modulo $d$, i.e. $n,k,m\in Z_d = \{0,1,...,d-1\}$. The odd value
of $d$ allows all independent indices to be covered in two cycles of $Z_d$.

This definition does not work for even $d$. If the index shift is made
one sided, the Wigner function does not remain real. So an alternative
construction is needed, incorporating a ``quantum square-root".
Since any integer is an odd number times a power of two, figuring out
the Wigner function for $d=2$ (i.e. a qubit) is sufficient to reach any
$d$ using tensor products.

For $d=2$, the Wigner function can be defined using eigenvalues of
$\sigma_z$ and $\sigma_x$ as the two conjugate labels (replacing $x$
and $p$). $\sigma_z$ and $\sigma_x$ are related by the Hadamard operator,
$\sigma_z = H\sigma_xH$, which gives the discrete Fourier transformation
in $d=2$. For instance, one can call $W(+,+)$ the weight for the spin
being up along both $z$-axis and $x$-axis. The Wigner function for a
qubit can be constructed as a map from the Bloch sphere representation,
$\rho=(I+\hat{n}\cdot\vec{\sigma})/2$, with the replacements:
\begin{eqnarray}
I ~\rightarrow~ \frac{1}{2}\begin{pmatrix} 1 & 1\cr 1 & 1 \end{pmatrix} &,&~~
\sigma_x ~\rightarrow~ \frac{1}{2}\begin{pmatrix} 1 & -1\cr 1 & -1 \end{pmatrix} ~,\cr
\sigma_y ~\rightarrow~ \pm\frac{1}{2}\begin{pmatrix} 1 & -1\cr -1 & 1 \end{pmatrix} &,&~~
\sigma_z ~\rightarrow~ \frac{1}{2}\begin{pmatrix} 1 & 1\cr -1 & -1 \end{pmatrix} ~.
\end{eqnarray}
The ambiguity in the sign for $\sigma_y$ is related to the
charge conjugation symmetry of the $SU(2)$ group algebra,
$\vec{\sigma}\leftrightarrow-\vec{\sigma^*}$,
and both choices should be checked for consistency.
The normalisation condition Tr$(\rho)=1$ becomes $\sum_{ij}W(i,j)=1$,
while $\rho^2\preceq\rho$ gives $\sum_{ij}W(i,j)^2 \le 1/2$.

The expectation values can be expressed as
$\langle O\rangle = \sum_{ij}W(i,j)~O(i,j)$,
where the operator normalisation, fixed by imposing $\langle I\rangle=1$,
is different from that for the Wigner function. The qubit operators map as:
\begin{equation}
I ~\rightarrow~ \begin{pmatrix} 1 & 1 \cr 1 & 1 \end{pmatrix} ~,~
\vec{\sigma}\cdot\hat{m} ~\rightarrow~
\begin{pmatrix} m_x \pm m_y + m_z ~&~ -m_x \mp m_y + m_z \cr
                m_x \mp m_y - m_z ~&~ -m_x \pm m_y - m_z \end{pmatrix} ~.
\end{equation}
The marginals giving qubit observables $\langle I\pm\sigma_i \rangle$
are all non-negative, while the expectation values are
$\langle\vec{\sigma}\cdot\hat{m}\rangle = \hat{n}\cdot\hat{m}$.

In this convention, the Wigner function is non-negative within the octahedron
$\pm x \pm y \pm z = 1$ embedded in the Bloch sphere (taking into account both
the signs of $\sigma_y$). The directions $\hat{n}_j \in \{\pm1,\pm1,\pm1\}$,
orthogonal to the faces of the octahedron, give the maximum negativity to the
Wigner function.

Wigner functions for multi-qubit states are easily constructed using tensor
products. For example, the density matrix for the two-qubit singlet state is:
\begin{equation}
\rho_{\rm singlet} = \frac{1}{4}(I\otimes I - \sigma_x\otimes\sigma_x
                   - \sigma_y\otimes\sigma_y - \sigma_z\otimes\sigma_z) ~,
\end{equation}
which, with the above replacements, yields the Wigner function:
\begin{equation}
W_{\rm singlet} = \frac{1}{8}\begin{pmatrix}
                  -1 & +1 & +1 & -1\cr +1 & +1 & +1 & +1\cr
                  +1 & +1 & +1 & +1\cr -1 & +1 & +1 & -1\end{pmatrix} ~.
\end{equation}
Its negative components are enough to give
$\langle(\vec{\sigma}\cdot\vec{a})(\vec{\sigma}\cdot\vec{b})\rangle
= -\vec{a}\cdot\vec{b}$, and violate the Bell inequality. Representing
$\vec{a}$ and $\vec{b}$ on the Bloch sphere, the positive and negative
hemispheres around them yield the measurement outcomes $+1$ and $-1$.
Then $\langle(\vec{\sigma}\cdot\vec{a})(\vec{\sigma}\cdot\vec{b})\rangle$
is the sum of four independent components with probabilities
$p_{++}=\frac{1}{4}(1-\vec{a}\cdot\vec{b})=p_{--}$,
$p_{+-}=\frac{1}{4}(1+\vec{a}\cdot\vec{b})=p_{-+}$.

\subsubsection{Quantum features}

Bell-type inequalities for experimentally observable correlations are derived
assuming statistical probability distributions for unobserved local hidden
variables. The standard formulation of quantum theory bypasses them, without
introducing any new variables, when the statistical probability distributions
are replaced by the density matrix. Quantum density matrices bring in complex
weights in general, Wigner functions make the weights real by a particular
choice of representation, but the possibility of the weights being
non-probabilistic (e.g. negative) remains. That is the sense in which Wigner
functions are different from classical phase space distributions.

Properties of Wigner functions have been useful in understanding how quantum
algorithms can be advantageous compared to their classical counterparts.
The Clifford group operations are those that transform the Pauli group
$\{I,\sigma_x,\sigma_y,\sigma_z\}^{\otimes n}$ within itself, up to phase
factors $\{\pm 1,\pm i\}$. For a single qubit, these operations are the
symmetry operations of the octahedron described in the previous section,
which transform the non-negative Wigner function region to itself. The
Gottesman-Knill theorem proves that all Clifford group operations can be
perfectly simulated in polynomial time on a probabilistic classical computer
\cite{gottesman}.

For a general quantum algorithm, Wigner functions can be associated with
the initial product state, the logic gate operations, and the final local
measurements. The outcome probabilities of any quantum algorithm can then
be expressed as a phase space probability distribution, which is a product
of these Wigner function factors summed over all evolution time steps $t$
and all quantum state components $n$. When all the Wigner function factors
are non-negative, the evolution describes a classical stochastic process,
which can be efficiently sampled with an effort polynomial in $n$ and $t$
\cite{eisert}. This result is robust with respect to sampling errors and
bounded approximations, and generalises the Gottesman-Knill theorem.

\subsection{Schmidt Decomposition}

This is a striking result from linear algebra, which predates quantum theory.
It simplifies the description of correlations between two complementary parts
of a quantum system, by making a clever choice of basis.

Any pure quantum state of a bipartite system can be expressed in the form:
\begin{equation}
\vert\psi_{AB}\rangle
= \sum_{i,\mu} a_{i\mu}\vert i_A\rangle\vert\mu_B\rangle
\equiv \sum_i \vert i_A\rangle\vert\overline{i}_B\rangle ~,
\end{equation}
where $\vert i_A\rangle\in{\cal H}_A$ and $\vert\mu_B\rangle\in{\cal H}_B$
form complete orthonormal bases, while the vectors
$\vert\overline{i}_B\rangle \equiv \sum_\mu a_{i\mu}\vert\mu_B\rangle \in {\cal H}_B$
may not be either normalised or mutually orthogonal. Now choose the
orthonormal basis $\{\vert i_A\rangle\}$ such that the reduced density
matrix $\rho_A$ is diagonal. $\rho_A$ can be also expressed as the partial
trace ${\rm Tr}_B(\rho_{AB})$. Comparison of the two forms gives:
\begin{eqnarray}
\rho_A &=& \sum_i p_i\vert i_A\rangle\langle i_A\vert \cr
&=& {\rm Tr}_B\Big( \big(\sum_i\vert i_A\rangle\vert\overline{i}_B\rangle\big)
    \big(\sum_j\langle j_A\vert\langle\overline{j}_B\vert\big) \Big)
= \sum_{ij} \langle\overline{j}_B\vert\overline{i}_B\rangle~\vert i_A\rangle\langle j_A\vert ~.
\end{eqnarray}
Consistency in the orthonormal basis $\{\vert i_A\rangle\}$ requires that
$\sum_j \langle\overline{j}_B\vert \overline{i}_B\rangle = p_i\delta_{ij}$.
Thus $\{\vert i_B\rangle\}$ also form an orthogonal basis, and the vectors
$\vert i_B'\rangle = p_i^{-1/2}\vert\overline{i}_B\rangle$ are orthonormal.
Moreover, we can also express
$\vert\psi_{AB}\rangle = \sum_i p_i^{1/2} \vert i_A\rangle\vert i_B'\rangle$,
and have the reduced density matrix
$\rho_B = \sum_i p_i\vert i_B'\rangle\langle i_B'\vert$.

This result, which converts a bipartite quantum state from a double sum over
indices $i$ and $\mu$ to a single sum over index $i$, by a clever choice of
basis, has many physical implications (subject to the specific choice of
partition):\\
$\bullet$
There is no restrictions on the dimensionalities of ${\cal H}_A$ and 
${\cal H}_B$. The number of non-zero values of $p_i$ that appear in the
preceding expansions of the reduced density matrices $\rho_A$ and $\rho_B$
is called the Schmidt rank $r_S$. Obviously,
$r_S \le {\rm min}({\rm dim}({\cal H}_A),{\rm dim}({\cal H_B}))$.
When $r_S=1$, the quantum state factorises between parts $A$ and $B$,
and there are no correlations. But when $r_S>1$, the quantum state does
not factorise between parts and $A$ and $B$, i.e. they are entangled.\\
$\bullet$
When ${\rm dim}({\cal H}_A)\le {\rm dim}({\cal H}_B)$, only up to
${\rm dim}({\cal H}_A)$ degrees of freedom of ${\cal H}_B$ can be correlated
with those of ${\cal H}_A$. This is true even if ${\cal H}_B$ has many more
degrees of freedom than ${\cal H}_A$, as is often the case when $A$ labels
the system and $B$ its environment. Diagonalisation of $\rho_B$ is needed to
explicitly find these degrees of freedom, but diagonalisation of $\rho_A$
is enough to specify their number. This one-to-one correspondence between
$\vert i_A\rangle$ and $\vert i_B'\rangle$ is a constraint on the bipartite
correlations, known as ``monogamy".\\
$\bullet$
The orthonormal basis sets $\{\vert i_A\rangle\}$ and $\{\vert i_B'\rangle\}$
with non-zero values of $p_i$ have the same size. So they can be related by
a unitary transformation (including both rotations and reflections). Also,
the Schmidt decomposition is unaffected by independent local unitary
transformations on the two parts. Any transformation of the form
$U_A\otimes U_B$ merely redefines the basis sets $\{\vert i_A\rangle\}$
and $\{\vert i_B'\rangle\}$.\\
$\bullet$
Since any mixed state density matrix can be diagonalised as
$\rho_A = \sum_i p_i\vert i_A\rangle\langle i_A\vert$, it can always be
extended to a pure state by adding suitable $\vert i_B'\rangle$. Such an
extension of a mixed state to a pure state is not unique, but the required
number of $\vert i_B'\rangle$ does not exceed dim$({\cal H}_A)$, and so the
pure state dimension does not exceed $({\rm dim}({\cal H}_A))^2$. This concept
turns out to be very useful in construction of error-correction codes for
bounded error quantum computation that eliminate undesired system-environment
correlations. It is also useful in construction of error mitigation schemes
that focus on removing the dominant errors corresponding to non-leading large
$p_i$.\\
$\bullet$
The correlations between the two parts of a pure quantum state can be
quantified in terms of the entropy:
\begin{equation}
S(\{p_i\}) = -\sum_i p_i \log(p_i)
= -{\rm Tr}(\rho_A\log(\rho_A)) = -{\rm Tr}(\rho_B\log(\rho_B)) ~.
\end{equation}
Noting that $S(\vert\psi_{AB}\rangle) = -{\rm Tr}(\rho_{AB}\log(\rho_{AB}))
= 0$ for the pure state $\vert\psi_{AB}\rangle$, $S(\{p_i\})$ is called the
{\em entropy of formation} of the mixed state. $S(\{p_i\})$ is maximised
when all $p_i$ are equal, $S_{\rm max}=\log(r_S)$. That corresponds to
equipartition or the microcanonical ensemble of statistical mechanics.\\
$\bullet$
For a system of two qubits, the Schmidt decomposition is
$\vert\psi_{AB}\rangle = \sqrt{p} \vert i_A\rangle\vert i_B'\rangle
                       + \sqrt{1-p} \vert j_A\rangle\vert j_B'\rangle$,
with $p\in[0,1]$ and $i\ne j$. In this case, the entropy $S(p)$ is a
monotonically increasing function of $p$ for $p\in[0,\frac{1}{2}]$, and
can be used to compare correlations between the two qubits, i.e. specify
whether one two-qubit system is more or less correlated than another one.
The choice $p=\frac{1}{2}$ gives the maximally entangled Bell states,
which form a complete orthonormal basis in the four-dimensional Hilbert
space. With the one-to-one correspondence between $\vert i_A\rangle$
and $\vert i_B'\rangle$, they are very useful in construction of quantum
cryptographic protocols.

\subsection{Quantum Statistics}

Quantum physics brings in the notion of identical particles, a concept
that is absent in classical physics. This feature produces many striking
effects, when more than one identical particles are present in a system.
With the inclusion of special relativity, the framework of quantum field
theory expands the scope of such telltale effects even further.

For evolution of $n$ identical particles, $n!$ permutation possibilities 
need to be summed over. Equivalently, there are $n!$ possible paths that
interfere giving rise to quantum effects. Some of the outstanding examples
are:\\
$\bullet$ The spin-statistics connection is a fundamental property that
relates the intrinsic spin of a particle and the statistics of collections
of such particles \cite{spinstatistics}. It can be understood geometrically
as the exchange of two identical particles being equivalent to rotating one
of them by angle $2\pi$. Its explicit proof requires relativistic quantum
field theory and representation theory of the Lorentz group, although it can
be simplified assuming the existence of antiparticles. Only two possibilities
occur in 3+1 dimensions: Symmetric bosons with integral spin and antisymmetric
fermions with half-integral spin, with two exchanges returning the quantum
state to its original form. On the other hand, any value of spin is allowed
for hard-core (quasi)particles in 2+1 dimensions.\\
$\bullet$ The symmetric Bose-Einstein statistics and antisymmetric Fermi-Dirac
statistics form the basis of quantum statistical physics. In reality, the
basic building blocks of matter are all fermions, while their interactions
are mediated by bosons. Beyond that, the statistical rules apply to composite
quantum objects as well. Anyonic statistics can be more complicated; it is
analysed using concepts of braiding and topology.\\
$\bullet$ Antisymmetric Fermi-Dirac statistics incorporates the Pauli
exclusion principle, i.e. any quantum state can contain either one fermion
or no fermion. That leads to the shell model for multi-electron atoms and
multi-nucleon nuclei as well as the Fermi sea and band structure of metals
and insulators.\\
$\bullet$ Degenerate Fermi gas produces pressure at its Fermi surface.
In situations where the temperature can be ignored, such as white dwarfs
and neutron stars, it gives their equation of state. When effects of
special relativity are included, the pressure has a limiting value.
When gravitational compression exceeds that pressure the stars collapse,
a dramatic consequence known as the Chandrasekhar limit.\\
$\bullet$ Unlike fermions, an unlimited number of bosons can occupy any
quantum state, e.g. as in case of lasers. Also, bosons can be produced
or absorbed singly, while fermions must be created or annihilated as
particle-antiparticle pairs. When the number of bosons is conserved, it is
possible for a macroscopic fraction of them to occupy the ground state of the
system at low enough temperatures, a phenomenon known as the Bose-Einstein
condensation where dissipation disappears. Condensation of hard-core bosons
produces superfluidity, while condensation of bound fermions pairs (which
are composite bosons) produces superconductivity. The complete lack of
dissipation in both cases is a distinctly quantum property.\\
$\bullet$ The Hong-Ou-Mandel effect demonstrates characteristic interference
between two identical photons \cite{hongoumandel}. When such photons enter
a 1:1 beam splitter simultaneously, one in each input port, they always exit
the beam splitter in the same output mode. There is zero probability for them
to exit separately as one each in the two output modes. This effect is useful
in accurately testing the distinguishability of the photons, in frequency
timing or path lengths.\\
$\bullet$ The Hanbury Brown and Twiss effect describes correlations in the
intensities received by two detectors from a narrow beam of particles
\cite{hanburybrowntwiss}. For a source of identical particles, there are
two possibilities for paths to the detectors, parallel and crossed. Their
interference leads to oscillating intensity correlations (in arrival times
at the detectors), corresponding to bunching for bosons, antibunching for
fermions, and no correlations for coherent sources (e.g. lasers). This effect
has been used to resolve the source sizes from radio astronomy to heavy-ion
collisions.

\section{Outlook}

Where does all this leave us? While it is certainly worthwhile to keep on
contemplating about foundational questions and interpretation of quantum
mechanics, it is best to follow Mermin's advice for practical applications
of quantum mechanics: ``Shut up and calculate!"
In fact, the Physics Breakthrough Prize for 2022 was awarded to Charles H.
Bennett, Gilles Brassard, David Deutsch and Peter Shor ``for foundational
work in the field of quantum information" \cite{PhyBreak22}. They have used
the well-established features of quantum theory to direct progress in the
rapidly growing field of quantum technology.
 
\bmhead{Acknowledgments}

This work was supported in part by the MSIL Chair Professorship at IISc.

\section*{Declarations}

\begin{itemize}
\item The corresponding author states that there is no conflict of interest.
\end{itemize}

\bibliography{sn-bibliography}


%

\end{document}